\documentclass[
floatfix,
twocolumn,
showpacs,showkeys,preprintnumbers,aps,prc]
{revtex4}

\usepackage{amsmath}
\usepackage{amssymb}
\usepackage{revsymb}
\usepackage{graphicx}
\usepackage{dcolumn}

\begin{document}

\title{Covariant calculation of strange decays of baryon resonances}
\author{B. Sengl, T. Melde, W. Plessas}
\affiliation{Theoretische Physik, Institut f\"ur Physik,
Karl-Franzens-Universit\"at \\
Universit\"atsplatz 5, A-8010 Graz, Austria}

\begin{abstract}
We present results for kaon decay widths of baryon resonances
from a relativistic study with constituent quark models.
The calculations are done in the point-form of Poincar\'e-invariant
quantum mechanics with a spectator-model decay operator.
We obtain covariant predictions of the Goldstone-boson-exchange and
a variant of the one-gluon-exchange constituent quark models for all
kaon decay widths of established baryon resonances.
They are generally characterized by underestimating the
available experimental data. In particular, the widths of kaon decays
with increasing strangeness in the baryon turn out to be extremely
small. We also consider the nonrelativistic limit, leading to the
familiar elementary emission model, and demonstrate
the importance of relativistic effects. It is found that the
nonrelativistic approach evidently misses sensible influences
from Lorentz boosts and some essential spin-coupling terms.  
\end{abstract}


\maketitle
\section{Introduction}
Recently, we studied the nonstrange decays of light baryons with relativistic
constituent quark models (CQMs)~\cite{Melde:2005hy}. The pertinent covariant
results for the partial widths calculated in point-form quantum mechanics
generally led to an underestimation of the experimental data. The
extension of the same type of investigations to nonstrange decays of
strange baryon resonances produced congruent properties of the corresponding
partial decay widths~\cite{Melde:2006zz}. As a result quite a consistent
behavior of the relativistic predictions emerged for all $\pi$ and $\eta$ 
decay widths~\cite{Melde:2006hb,Sengl:2006gz}. 
Previous investigations~\cite{Koniuk:1980vy,LeYaouanc:1988aa,
Stancu:1988gb,Stancu:1989iu,Capstick:1993th,Capstick:1994kb,
Geiger:1994kr,Ackleh:1996yt,Plessas:1999nb,Theussl:2000sj}
of these decays were performed mostly
along nonrelativistic or relativized approaches and with different dynamical
models and/or various decay operators. They yielded
strongly varying results, and the employment of various parametrizations
in the decay models made it difficult to compare them with each other.

While the nonstrange decays of baryons have received a fair amount of attention,
CQM investigations of strange decays have remained rather limited. Koniuk and
Isgur~\cite{Koniuk:1980vy} made a first study of $K$ decays along the
nonrelativistic elementary emission model (EEM) employing
a one-gluon-exchange (OGE) CQM. Later on, Capstick and
Roberts~\cite{Capstick:1998uh} reported results for
strange decays of nonstrange baryons with a quark-pair-creation operator
using a relativized OGE CQM. In any case, the theoretical description of
strange baryon decays lacks behind phenomenology, as there are a number of
well established data reported from experiments (see, e.g., the compilation
by the Particle Data Group (PDG)~\cite{PDBook}).

Here, we extend our relativistic studies of $\pi$ and $\eta$ decays of light
and strange baryon resonances~\cite{Melde:2005hy,Melde:2006zz} to the $K$
channels. We report covariant predictions for the $K$ decay widths of the
established $N$, $\Delta$, $\Lambda$, $\Sigma$, and $\Xi$ resonances below
2 GeV. We employ two
types of dynamical models, namely, the Goldstone-boson-exchange (GBE)
CQM~\cite{Glozman:1998ag,Glozman:1998fs} and a relativistic variant of the
Bhaduri-Cohler-Nogami OGE CQM~\cite{Bhaduri:1980fd}
as parametrized in Ref.~\cite{Theussl:2000sj}. The calculations are
performed in the framework of Poincar\'e-invariant quantum
mechanics~\cite{Keister:1991sb} in point form~\cite{Dirac:1949}. In
particular, we employ the decay operator from the so-called point-form
spectator model (PFSM)~\cite{Melde:2005hy,Melde:2004qu,Melde:2006zz} producing
frame independent results for the partial decay widths. We also consider
the nonrelativistic limit of the PFSM leading to the decay operator of the
usual EEM.
\section{Formalism}
The theory of the relativistic point-form treatment of mesonic decays can be
found in our earlier papers~\cite{Melde:2005hy,Melde:2006zz}, and we follow the
same notation. Here, we only give the fundamental formulae generalized to the
case of a decay operator changing the strangeness contents from the incoming
to the outgoing baryons. The corresponding decay width is given by
\begin{multline}
\label{eq:decwidth}
{\mit\Gamma}_{i\to f}=\frac{|{\vec q}|}{4M^2}\;\frac{1}{2J+1}
\\ \times
\sum\limits_{M_J,M_{J'}}
\frac{1}{2T+1}\sum\limits_{\;M_T,M_{T'},M_{T_m}}
|F_{i\to f}|^2 \, ,
\end{multline}
where $F_{i\to f}$ is the transition amplitude. The latter is defined by
the matrix element of the four-momentum conserving reduced decay
operator $\hat D^m_{rd}$ between incoming and outgoing baryon states
\begin{multline}
\label{eq:transel}
F_{i\to f}\\
=\langle V',M',J',M_{J'},T',M_{T'}|{\hat D}^m_{rd} |V,M,J,M_J,T,M_T \rangle
\, .
\end{multline}
Here, $q_\mu=(q_0,\vec q)$ denotes the four-momentum of the outgoing $K$ meson.
The invariant mass eigenstate of the incoming baryon resonance
$|V,M,J,M_J,T,M_T \rangle$ is characterized by the
eigenvalues of the velocity $V$, mass $M$, intrinsic spin $J$ with z-component
$M_J$, and isospin $T$ with z-projection $M_T$; correspondingly the
outgoing baryon eigenstate is denoted by primed eigenvalues. The matrix
element in Eq.~(\ref{eq:transel}) can be expressed in an appropriate
basis representation using velocity states through the integral (for details
of the calculation see the appendix of Ref.~\cite{Melde:2005hy})
\begin{widetext}%
\begin{eqnarray}
&&
    {\langle V',M',J',M_{J'},T',M_{T'}|{\hat D}_{rd}^m|
    V,M,J,M_J,T,M_T\rangle}
    =
     {\frac{2}{MM'}\sum_{\sigma_i\sigma'_i}\sum_{\mu_i\mu'_i}{
    \int{
    d^3{\vec k}_2d^3{\vec k}_3d^3{\vec k}'_2d^3{\vec k}'_3
    }} }
\nonumber\\
&&
{\times \sqrt{\frac{\left(\sum_i \omega'_i\right)^3}
    {\prod_i 2\omega'_i}}
    \Psi^\star_{M'J'M_{J'}T'M_{T'}}\left({\vec k}'_1,{\vec k}'_2,{\vec k}'_3;
    \mu'_1,\mu'_2,\mu'_3\right)
    \prod_{\sigma'_i}{D_{\sigma'_i\mu'_i}^{\star \frac{1}{2}}
    \left\{R_W\left[k'_i;B\left(V'\right)\right]\right\}
    }
    }
\nonumber\\
&&
     {\times
     \left<p'_1,p'_2,p'_3;\sigma'_1,\sigma'_2,\sigma'_3\right|{\hat D}_{rd}^m
    \left|p_1,p_2,p_3;\sigma_1,\sigma_2,\sigma_3\right>
    \nonumber }
\nonumber\\
&&
    \times \prod_{\sigma_i}{D_{\sigma_i\mu_i}^{\frac{1}{2}}
    \left\{R_W\left[k_i;B\left(V\right)\right]\right\}}
         {\sqrt{\frac{\left(\sum_i \omega_i\right)^3}
    {\prod_i 2\omega_i}}
    \Psi_{MJM_J TM_T}\left({\vec k}_1,{\vec k}_2,{\vec k}_3;\mu_1,
    \mu_2,\mu_3\right)} \, ,
\end{eqnarray}
\end{widetext}%
where $\Psi_{MJM_J TM_T}\left({\vec k}_1,{\vec k}_2,{\vec
k}_3;\mu_1,\mu_2,\mu_3\right) $ and
$\Psi^\star_{M'J'M_{J'}T'M_{T'}}\left({\vec k}'_1,{\vec k}'_2,{\vec
k}'_3;\mu'_1,\mu'_2,\mu'_3\right)$ are the rest-frame wave functions
of the incoming and outgoing baryons, respectively. The
three-momenta $\vec k_i$ satisfy the rest-frame condition
$\sum_i{\vec k_i}=\vec 0$. They are connected to the four-momenta
$p_i$ by the boost relations $p_i=B\left(V\right)k_i$. The Wigner
$D$-functions $D_{\sigma_i\mu_i}^{\frac{1}{2}}$, describing the effects
of the Lorentz boosts on the internal spin states, follow directly from
the representation of the baryon eigenstates with velocity states.
The particular form of the decay operator is taken according to the
PFSM construction~\cite{Melde:2004qu} and reads as
\begin{multline}
\label{momrepresent}
\langle p'_1,p'_2,p'_3;\sigma'_1,\sigma'_2,\sigma'_3
|{\hat D}^{m}_{rd}|
 p_1,p_2,p_3;\sigma_1,\sigma_2,\sigma_3\rangle
 =
 \\-3{\left(\frac{M}{\sum_i{\omega_i}}
\frac{M'}{\sum_i{\omega'_i}}\right)^{\frac{3}{2}} }
\frac{ig_{qqm}}{m_1+m'_1}\frac{1}{\sqrt{2\pi}}
\\
\times \bar{u}(p_1',\sigma_1')\gamma_5\gamma^\mu \mathcal{F}^m
 u(p_1,\sigma_1)q_\mu
 \\
\times 2p_{20}\delta^3\left(
{\vec {p}}_2-{\vec {p}}'_2\right)\delta_{\sigma_2 \sigma'_2}
2p_{30}\delta^3\left({\vec {p}}_3-{\vec {p}}'_3\right)
\delta_{\sigma_3 \sigma'_3} \, .
\end{multline}
Here, $g_{qqm}$ is the quark-meson coupling constant and $\mathcal{F}^m$ the
flavor-transition operator specifying the particular decay mode. The masses
$m_1$ and $m'_1$ refer to the active quark in the incoming and outgoing channels,
respectively. More details on the formalism can be found also in
Ref.~\cite{Sengl:2006}.

In order to demonstrate the relativistic effects we also consider the
nonrelativistic limit of the PFSM. It corresponds to a decay operator
along the EEM, i.e.:
\begin{multline}
\label{eq:NR12}
F_{i\to f}^{NR}
=\sqrt{2E}\sqrt{2E'}\sum_{\mu_i\mu'_i}
 \int{d^3k_2d^3k_3}
\\
\times \Psi^\star_{M'J'M_{J'}T'M_{T'}}\left({\vec k}'_1,{\vec k}'_2,{\vec k}'_3;
    \mu'_1,\mu'_2,\mu'_3\right)
\\
 \times \frac{-3ig_{qqm}}{m_1+m'_1} \frac{1}{\sqrt{2\pi}}\mathcal{F}^m
 \left\{\left[
 -\omega_m\frac{\left(m_1+m'_1\right)}{2m_1m'_1}
\vec \sigma_1 \cdot {\vec{p}}_1
\right.\right.
\\
\left.\left.
+\left(1+\frac{\omega_m}{2m'_1}\right)
\vec \sigma_1 \cdot {\vec{q}}
\right]
\delta_{\mu_2\mu'_2}\delta_{\mu_3\mu'_3}
\right\}_{\mu'_i\mu_i}
\\
\times \Psi_{MJM_J TM_T}\left({\vec k}_1,{\vec k}_2,{\vec k}_3;\mu_1,
    \mu_2,\mu_3\right)\;.
\end{multline}
A derivation of this limit can be found in Ref.~\cite{Melde:2006zz}. Here
we give the generalized result including the case with the masses $m_1$ and
$m'_1$ being different.

We note that such a nonrelativistic study of strange decays in the context of the
EEM was already performed before, reporting in particular predictions of the GBE
CQM~\cite{Krassnigg:1999ky}. However, one employed an even more simplified
nonrelativistic decay operator neglecting the mass differences of the
active quark in the incoming and outgoing channels. Consequently, we may
consider the present nonrelativistic results, quoted in the Tables below,
to supersede the ones reported in Ref.~\cite{Krassnigg:1999ky}.

\section{Results}

In our study we consider the baryon resonances contained in
Table~\ref{tab:masses}. The partial $K$ decay widths are calculated using
directly the wave functions as produced by the GBE CQM and
the Bhaduri-Cohler-Nogami OGE CQM, where in
one calculation theoretical baryon masses and in another calculation
experimental ones are used as inputs. The decay channels
with decreasing strangeness in the baryon are contained in Table~\ref{K_s_to_u},
the ones with increasing strangeness in Table~\ref{K_u_to_s}. In these tables
the first and second columns always denote the decaying resonances with their
intrinsic spins and parities $J^P$, and the third columns give the experimental
decay widths with their uncertainties, as quoted by the
PDG~\cite{PDBook}.

\renewcommand{\arraystretch}{1.5}
\begin{table}[h!]
\begin{center}
\caption{Theoretical and experimental masses (in MeV) of the ground and resonance
states considered in the decay calculations.
\label{tab:masses}
}
\begin{ruledtabular}
\begin{tabular}{ccccc}
Baryon&$J^P$&\multicolumn{2}{c}{Theory}&Experiment\\
&&GBE&OGE&\cite{PDBook} \\
\hline
$N$
&$\frac{1}{2}^+$
&$939$
&$939$
&$938 -940$
\\
$N(1440)$
&$\frac{1}{2}^+$
&$1459$
&$1577$
&$1420-1470$
\\
$N(1520)$
&$\frac{3}{2}^-$
&$1519$
&$1521$
&$1515-1525$
\\
$N(1535)$
&$\frac{1}{2}^-$
&$1519$
&$1521$
&$1525-1545$
\\
$N(1650)$
&$\frac{1}{2}^-$
&$1647$
&$1690$
&$1645-1670$
\\
$N(1675)$
&$\frac{5}{2}^-$
&$1647$
&$1690$
&$1670-1680$
\\
$N(1700)$
&$\frac{3}{2}^-$
&$1647$
&$1690$
&$1650-1750$
\\
$N(1710)$
&$\frac{1}{2}^+$
&$1776$
&$1859$
&$1680-1740$
\\
\hline
$\Delta$
&$\frac{3}{2}^+$
&$1240$
&$1231$
&$1231-1233$
\\
$\Delta(1600)$
&$\frac{3}{2}^+$
&$1718$
&$1854$
&$1550-1700$
\\
$\Delta(1620)$
&$\frac{1}{2}^-$
&$1642$
&$1621$
&$1600-1660$
\\
$\Delta(1700)$
&$\frac{3}{2}^-$
&$1642$
&$1621$
&$1670-1750$
\\
\hline
$\Lambda$
&$\frac{1}{2}^+$
&$1136$
&$1113$
&$1116$
\\
$\Lambda(1405)$
&$\frac{1}{2}^-$
&$1556$
&$1628$
&$1402-1410$
\\
$\Lambda(1520)$
&$\frac{3}{2}^-$
&$1556$
&$1628$
&$1519-1521$
\\
$\Lambda(1600)$
&$\frac{1}{2}^+$
&$1625$
&$1747$
&$1560-1700$
\\
$\Lambda(1670)$
&$\frac{1}{2}^-$
&$1682$
&$1734$
&$1660-1680$
\\
$\Lambda(1690)$
&$\frac{3}{2}^-$
&$1682$
&$1734$
&$1685-1695$
\\
$\Lambda(1800)$
&$\frac{1}{2}^-$
&$1778$
&$1844$
&$1720-1850$
\\
$\Lambda(1810)$
&$\frac{1}{2}^+$
&$1799$
&$1957$
&$1750-1850$
\\
$\Lambda(1830)$
&$\frac{5}{2}^-$
&$1778$
&$1844$
&$1810-1830$
\\
\hline
$\Sigma$
&$\frac{1}{2}^+$
&$1180$
&$1213$
&$1189-1197$
\\
$\Sigma(1660)$
&$\frac{1}{2}^+$
&$1616$
&$1845$
&$1630-1690$
\\
$\Sigma(1670)$
&$\frac{3}{2}^-$
&$1677$
&$1732$
&$1665-1685$
\\
$\Sigma(1750)$
&$\frac{1}{2}^-$
&$1759$
&$1784$
&$1730-1800$
\\
$\Sigma(1775)$
&$\frac{5}{2}^-$
&$1736$
&$1829$
&$1770-1780$
\\
$\Sigma(1880)$
&$\frac{1}{2}^+$
&$1911$
&$2049$
&$1806-2025$
\\
$\Sigma(1940)$
&$\frac{3}{2}^-$
&$1736$
&$1829$
&$1900-1950$
\\
\hline
$\Xi$
&$\frac{1}{2}^+$
&$1348$
&$1346$
&$1315-1321$
\\
$\Xi(1820)$
&$\frac{3}{2}^-$
&$1792$
&$1894$
&$1818-1828$
\\
\end{tabular}
\end{ruledtabular}
\end{center}
\end{table}

\renewcommand{\arraystretch}{1.5}
\begin{table*}
\caption{Covariant predictions by the GBE CQM~\cite{Glozman:1998ag} and
the OGE CQM~\cite{Theussl:2000sj} for $K$ decay widths with decreasing
strangeness in the baryon. Experimental data as reported by the
PDG~\cite{PDBook}. Both theoretical and experimental masses are used as input.
In the last three columns a comparison is given to results
by Feynman, Kislinger, Ravndal (FKR)~\cite{Feynman:1971wr},
Koniuk and Isgur (KI)~\cite{Koniuk:1980vy} as well as Capstick and Roberts
(CR)~\cite{Capstick:1998uh}. Regarding the FKR result for $\Lambda$(1810) the
$\star$ indicates that this state was attributed a different intrinsic spin
of $J^P=\frac{5}{2}^+$.
\label{K_s_to_u}} \vspace{0.2cm}
\begin{ruledtabular}
{\begin{tabular}{@{}ccr|c cc c|c ccc|cc@{}}
&&
&\multicolumn{4}{c}{With Theoretical Mass}
&\multicolumn{4}{c}{With Experimental Mass}
&\multicolumn{2}{c}{}
\\
Decay &$J^P$&Experiment \small [MeV]
&\multicolumn{2}{c}{Relativistic}
&\multicolumn{2}{c|}{Nonrel. EEM}
&\multicolumn{2}{c}{Relativistic}
&\multicolumn{2}{c|}{Nonrel. EEM}
&\multicolumn{2}{c}{Literature}
\\
\small $\rightarrow NK $&&& GBE & OGE & GBE & OGE & GBE & OGE&GBE & OGE & FKR & KI \\
\hline
$\Lambda(1520)$&$\frac{3}{2}^-$
    &$\left(7.02\pm0.16\right)_{-0.44}^{+0.46}$
    &$12$
    &$24$
    &$23$
    &$63$
    &$6$
    &$5$
    &$13$
    &$19$
    &$7$
    &$9.0$
\\
$\Lambda(1600)$&$\frac{1}{2}^+$
    &$\left(33.75\pm11.25\right)_{-15}^{+30}$
    &$15$
    &$35$
    &$4.1$
    &$23$
    &$14$
    &$21$
    &$3.8$
    &$11$
    &$$
    &$29$
\\
$\Lambda(1670)$&$\frac{1}{2}^-$
    &$\left(8.75\pm1.75\right)_{-\phantom{0}2}^{+4.5}$
    &$0.3$
    &$\approx 0$
    &$45$
    &$86$
    &$0.4$
    &$0.4$
    &$45$
    &$76$
    &$415$
    &$11$
\\
$\Lambda(1690)$&$\frac{3}{2}^-$
    &$\left(15\pm3\right)_{-2}^{+3}$
    &$1.2$
    &$1.0$
    &$4.2$
    &$6.5$
    &$1.2$
    &$0.8$
    &$4.5$
    &$4.7$
    &$102$
    &$15$
\\
$\Lambda(1800)$&$\frac{1}{2}^-$
    &$\left(97.5\pm22.5\right)_{-25}^{+40}$
    &$4.2$
    &$6.4$
    &$3.1$
    &$8.6$
    &$4.5$
    &$5.5$
    &$3.3$
    &$7.4$
    &$$
    &$8.4$
\\
$\Lambda(1810)$&$\frac{1}{2}^+$
    &$\left(52.5\pm22.5\right)_{-20}^{+50}$
    &$4.1$
    &$12$
    &$23$
    &$44$
    &$4.8$
    &$2.2$
    &$24$
    &$31$
    &$35^*$
    &$33$
\\
$\Lambda(1830)$&$\frac{5}{2}^-$
    &$\left(6.18\pm3.33\right)_{-1.05}^{+1.05}$
    &$0.1$
    &$0.9$
    &$0.1$
    &$0.1$
    &$1.2$
    &$0.5$
    &$0.2$
    &$0.1$
    &$0$
    &$2.3$
\\
$\Sigma(1660)$&$\frac{1}{2}^+$
    &$\left(20\pm10\right)_{-\phantom{0}6}^{+30}$
    &$0.9$
    &$0.9$
    &$0.4$
    &$\approx 0$
    &$1.2$
    &$0.5$
    &$0.5$
    &$0.02$
    &$$
    &$1.4$
\\
$\Sigma(1670)$&$\frac{3}{2}^-$
    &$\left(6.0\pm1.8\right)_{-1.4}^{+2.6}$
    &$1.1$
    &$1.0$
    &$1.9$
    &$2.0$
    &$1.1$
    &$0.6$
    &$1.8$
    &$1.3$
    &$3$
    &$4.4$
\\
$\Sigma(1750)$&$\frac{1}{2}^-$
    &$\left(22.5\pm13.5\right)_{-\phantom{0}3}^{+28}$
    &$\approx0$
    &$1.4$
    &$10$
    &$48$
    &$\approx0$
    &$0.8$
    &$10$
    &$45$
    &$14$
    &$17$
\\
$\Sigma(1775)$&$\frac{5}{2}^-$
    &$\left(48.0\pm3.6\right)_{-5.6}^{+6.5}$
    &$11$
    &$15$
    &$20$
    &$41$
    &$13$
    &$12$
    &$26$
    &$30$
    &$66$
    &$45$
\\
$\Sigma(1880)$&$\frac{1}{2}^+$
    &$$
    &$\approx 0$
    &$\approx0$
    &$5.4$
    &$13$
    &$\approx 0$
    &$0.1$
    &$5.1$
    &$8.1$
    &$$
    &$3.2$
\\
$\Sigma(1940)$&$\frac{3}{2}^-$
    &$\left(22\pm22\right)_{}^{+16}$
    &$1.1$
    &$1.5$
    &$3.3$
    &$6.8$
    &$2.3$
    &$2.1$
    &$10$
    &$12$
    &$$
    &$19$
\\
\hline
\multicolumn{1}{c}{\small $\rightarrow \Lambda K $}&
\multicolumn{12}{l}{}\\
\hline
$\Xi(1820)$&$\frac{3}{2}^-$
    &$large$
    &$2.7$
    &$6.2$
    &$6$
    &$4.5$
    &$3.5$
    &$19$
    &$10$
    &$11$
    &$15$
    &$$
\\
\hline \multicolumn{1}{c}{\small $\rightarrow \Sigma K $}&
\multicolumn{12}{l}{}\\
\hline
$\Xi(1820)$&$\frac{3}{2}^-$
    &$small$
    &$4.1$
    &$9.3$
    &$10$
    &$31$
    &$5.1$
    &$4.7$
    &$12$
    &$17$
    &$17$
    &$$
\end{tabular}}
\end{ruledtabular}
\end{table*}
\begin{table*}
\caption{Same as Table~\ref{K_s_to_u} but for $K$ decays with increasing
strangeness in the baryon.
\label{K_u_to_s}} \vspace{0.2cm}
\begin{ruledtabular}
{\begin{tabular}{@{}ccr| cccc| cccc| ccc@{}}
&&
&\multicolumn{4}{c}{With Theoretical Mass}
&\multicolumn{4}{c}{With Experimental Mass}
&\multicolumn{3}{c}{}\\
Decay &$J^P$&Experiment \small [MeV]
&\multicolumn{2}{c}{Relativistic}
&\multicolumn{2}{c|}{Nonrel. EEM}
&\multicolumn{2}{c}{Relativistic}
&\multicolumn{2}{c|}{Nonrel. EEM}
&\multicolumn{3}{c}{Literature}
\\
&&& GBE & OGE & GBE & OGE & GBE & OGE & GBE & OGE &FKR & KI & CR\\
\hline
\multicolumn{1}{c}{\small $\rightarrow \Lambda K $}&
\multicolumn{13}{l}{}\\
\hline
$N(1650)$&$\frac{1}{2}^-$
    &$\left(11.6\pm 6.6\right)_{-0.6}^{+2.2}$
    &$\approx 0$
    &$0.1$
    &$\approx 0$
    &$0.1$
    &$0.1$
    &$\approx 0$
    &$0.1$
    &$\approx 0$
    &$0$
    &$9$
    &$27$
\\
$N(1675)$&$\frac{5}{2}^-$
    &$<1.7$
    &$\approx 0$
    &$\approx 0$
    &$\approx 0$
    &$\approx 0$
    &$\approx 0$
    &$\approx 0$
    &$\approx 0$
    &$\approx 0$
    &$0$
    &$0.01$
    &$0$
\\
$N(1700)$&$\frac{3}{2}^-$
    &$\left(1.5\pm 1.5\right)_{}^{+1.5}$
    &$\approx 0$
    &$0.1$
    &$\approx 0$
    &$\approx 0$
    &$0.1$
    &$0.1$
    &$\approx 0$
    &$\approx 0$
    &$$
    &$0.04$
    &$0.16$
\\
$N(1710)$&$\frac{1}{2}^+$
    &$\left(15\pm 10\right)_{-\phantom{0}2.5}^{+37.5}$
    &$\approx 0$
    &$\approx 0$
    &$0.1$
    &$\approx 0$
    &$\approx 0$
    &$\approx 0$
    &$\approx 0$
    &$\approx 0$
    &$$
    &$4.4$
    &$7.8$
\\
\hline
\multicolumn{1}{c}{\small $\rightarrow \Sigma K $}&
\multicolumn{6}{l}{}\\
\hline
$N(1700)$&$\frac{3}{2}^-$
    &$$
    &$$
    &$$
    &$$
    &$$
    &$\approx 0$
    &$\approx 0$
    &$\approx 0$
    &$0.1$
    &$$
    &$small$
    &$0$
\\
$N(1710)$&$\frac{1}{2}^+$
    &$$
    &$3.9$
    &$389$
    &$2.8$
    &$739$
    &$0.4$
    &$27$
    &$0.3$
    &$42$
    &$$
    &$0.16$
    &$0.01$
\\
$\Delta(1700)$&$\frac{3}{2}^-$
    &$$
    &$$
    &$$
    &$$
    &$$
    &$\approx 0$
    &$\approx 0$
    &$\approx 0$
    &$0.1$
    &$$
    &$0.16$
    &$0.01$
\\
\hline
\multicolumn{1}{c}{\small $\rightarrow \Xi K $}&
\multicolumn{6}{l}{}\\
\hline
$\Lambda(1830)$&$\frac{5}{2}^-$
    &$$
    &$$
    &$\approx 0$
    &$$
    &$\approx 0$
    &$\approx 0$
    &$\approx 0$
    &$0.1$
    &$0.5$
    &$$
    &$$
    &$$
\\
$\Sigma(1880)$&$\frac{1}{2}^+$
    &$$
    &$0.1$
    &$0.1$
    &$\approx 0$
    &$\approx 0$
    &$0.1$
    &$0.2$
    &$\approx 0$
    &$\approx 0$
    &$$
    &$$
    &$$
\\
$\Sigma(1940)$&$\frac{3}{2}^-$
    &$$
    &$$
    &$$
    &$$
    &$$
    &$\approx 0$
    &$\approx 0$
    &$\approx 0$
    &$\approx 0$
    &$$
    &$$
    &$$
\end{tabular}}
\end{ruledtabular}
\end{table*}

The covariant CQM predictions obtained for the $K$ decay widths confirm the
picture that has already emerged with the relativistic results for the $\pi$
and $\eta$ decays of both the nonstrange and strange baryon
resonances~\cite{Melde:2005hy,Melde:2006zz}. The theoretical values are
usually smaller than the experimental data or at most reach them from below.
The only exception appears to be the $\Lambda(1520) \to NK$ decay in the first
entry of Table~\ref{K_s_to_u}. The relatively large decay widths found there,
however, are mainly due to the theoretical resonance masses reproduced too large
by the CQMs (cf. the values quoted in Table~\ref{tab:masses}). The decay widths
calculated with the experimental mass of 1520 MeV also remain below experiment.

The differences between the direct predictions of the GBE and OGE CQMs
are caused by two effects, the different theoretical baryon masses and the
different wave functions. In the calculations with experimental masses as
input the former (mass) effect is wiped out in the comparison of the two
CQMs. The corresponding results only
reflect the variations in the CQM wave functions. While the general
behavior of the decay widths is similar for both types of CQMs, we
nevertheless find some differences from the dynamics in certain decay
channels.

In the last two rows of Table~\ref{K_s_to_u} we also give the $K$ decay
widths of the $\Xi$(1820) resonance. In view of a total width of about 24 MeV
they are reported by the PDG~\cite{PDBook}
to be 'large' and 'small' for the $\Lambda K$ and
$\Sigma K$ channels, respectively. The relative sizes of the corresponding
partial decay widths are opposite in the theoretical CQM predictions.
Only, in the case of the OGE CQM, when the theoretical mass is replaced by
the experimental one, the
$K$ decay width of $\Xi(1820) \to \Lambda K$ becomes larger than the one
of $\Xi(1820) \to \Sigma K$. This behavior, however, is only found for
the relativistic PFSM and not in the nonrelativistic limit.

The CQM predictions for the decays with increasing strangeness in the baryon
shown in Table~\ref{K_u_to_s} are extremely small, practically vanishing;
there, figures smaller than 0.1 MeV are quoted as approximately zero.
However, there is one striking exception, namely, the decay 
$N(1710) \to \Sigma K$. Obviously, the corresponding results are dominated
by the phase-space factor. When removing the effects from the theoretical
masses, the decay widths are much reduced, even though a sensible effect
remains to be evident from the wave function in case of the OGE CQM. 
We note that the threshold for this $K$ decay channel is rather close to
the (central) mass value of the $N(1710)$ resonance. Consequently, also this
particular decay width is expected to be rather small, as it can
even vanish within the experimental bounds of the involved masses.
With regard to the decays with increasing strangeness in the baryon the PDG
quotes data only for four channels. Unfortunately, the uncertainties are
still rather large, and the present data could well allow for rather small
phenomenological decay widths, in line with the tendency of our results.

In Tables~\ref{K_s_to_u} and~\ref{K_u_to_s} we also quote the nonrelativistic
limit of our results; they correspond to the usual EEM. The
reduction of the PFSM consists in neglecting the Lorentz boosts and a
nonrelativistic expansion of the transition amplitude (for details
see Ref.~\cite{Melde:2006zz}). As has already been found before,
this implies a truncation of certain spin-coupling terms,
what may cause considerable effects in some decay channels. Similar findings
are made with the $K$ decays. The biggest changes are seen in the cases of
$\Lambda$(1670), $\Sigma$(1750), and $\Sigma$(1880) decays. While the
PFSM results are practically vanishing, the nonrelativistic decay widths
acquire respectable magnitudes.
Sizable enhancements furthermore occur for the  
$J^P=\frac{3}{2}^-$ resonances. The nonrelativistic reduction also causes
the predictions from the two CQMs to compare differently to each other than
is observed in the relativistic PFSM results. 

In the last columns of Tables~\ref{K_s_to_u} and~\ref{K_u_to_s} we quote
for comparison results from $K$ decay investigations by Feynman,
Kislinger, and Ravndal~\cite{Feynman:1971wr} (FKR), Koniuk and
Isgur~\cite{Koniuk:1980vy} (KI) as well as Capstick and
Roberts~\cite{Capstick:1998uh} (CR). While the first study is different
in nature, the latter two are done with CQMs. The calculations by
KI are essentially nonrelativistic using a decay operator of the EEM type.
In the work of CR a relativistic OGE CQM was employed and the decay
operator was taken according to the $^3P_0$ model. In both of the latter
works additional parametrisations of the decay amplitudes were introduced
to explore a possible fitting of the data. Consequently, it is not
surprising that in most cases an obvious agreement with experimental data
is reached.

Some useful insights can be gained from the comparison with the FKR results
even though much care has to be taken. The $K$ decay widths reported by
FKR are generally also rather small with the exception of a few cases.
Notable are the extremely large decay widths of $\Lambda(1670)$ and
$\Lambda(1690)$ going to $NK$. Both
of these results were questioned already by the authors themselves and for
the $\Lambda$(1670) a delicate cancellation among spin-coupling terms was
observed. We find a similar behavior in this decay channel. In
the fully relativistic PFSM calculation there occurs a nearly complete
cancellation of spin-coupling terms, while the truncation in the
EEM case picks up large contributions from terms surviving
the nonrelativistic reduction. On the other hand, FKR have not seen a
similar reason in case of $\Lambda$(1690). This is again congruent with
our observation that effects of truncated spin-coupling terms are not
prevailing in this channel. Regarding the FKR result reported for
$\Lambda$(1810) one should note that FKR attributed a $J^P$ different
from the established $\frac{1}{2}^+$. Therefore this entry must be
considered as obsolete. For the
$K$ decays with increasing strangeness in the baryon (Table~\ref{K_u_to_s})
FKR give two zero results quite similar to the extremely small (almost
vanishing) predictions we get. 

\section{Summary and Conclusions}
We have presented first covariant results for strange decays of baryon
resonances in the quark-model approach. In particular, we have given
predictions of the relativistic GBE and OGE CQMs of
Refs.~\cite{Glozman:1998ag} and~\cite{Theussl:2000sj},
respectively, for $K$ decay widths
calculated with the PFSM transition operator. It has been found that
the theoretical results obtained in this approach generally underestimate
the existing experimental data and are thus in line with the findings
made in preceding calculations of $\pi$ and $\eta$
decays~\cite{Melde:2005hy,Melde:2006zz}.
As has become evident by the comparison to the nonrelativistic limit
of the PFSM, which leads to the EEM often used in previous
investigations, relativistic effects play a decisive role.
Especially boost effects and all relativistic spin-coupling terms have
to be included in order to arrive at reliable theoretical results.

When comparing the class of $K$ decays with decreasing strangeness in
the baryon to the one with increasing strangeness, one observes a
striking distinction. All decay widths of the latter turn out to
be extremely small and may basically be considered as zero.
In any case, sizable $K$ decay widths are only found when the
strangeness decreases from the decaying resonance to the final
baryon.

In general, the CQMs with two different kind of dynamics lead to congruent
results. In a majority of cases the predictions of the GBE and OGE CQMs
produce rather similar predictions, at least when effects from different
theoretical resonance masses are removed, i.e. in the calculations with
experimental masses used as input. In some cases, however, also noticeable
influences from different CQM wave functions are found. Striking examples
are the $\Lambda(1810) \to NK$ or the $\Xi(1820) \to \Lambda K$ decays.
Of particular interest may also be the $N(1710) \to \Sigma K$ decay, since
the magnitude of the corresponding decay width is largely governed by
threshold effects. It might be advised to focus the attention of future
experiments to such decay channels with a sensible dependence on
different quark-quark dynamics.

In summary quite a consistent picture emerges for both the strange and
nonstrange decays. From the present approach one obtains CQM predictions
for decay widths that remain in general smaller than the experimentally
measured ones. In some cases the experimental data are at most reached
from below. These observations apply to all of our results for $\pi$, $\eta$,
and $K$ decays of nonstrange and strange baryons considered so far along
the relativistic PFSM. By finding essentially all decay widths being too
small a consistent deficiency becomes apparent that resides
either in the underlying CQMs and/or in the applied decay mechanism.
As a consequence one may not yet explain the phenomenology of the mesonic
decays of baryon resonances following the present approach. Still, we
consider establishing these results to be necessary benchmarks for future
studies. As we have not introduced any additional parametrizations beyond
the direct CQM predictions, they provide a fundamental basis for
further refinements.
As a most promising approach we consider the inclusion of additional mesonic
degrees of freedom directly on the baryon level (e.g., similar to the
method developed in Ref.~\cite{Canton:2000zf}).
For the improvement of the decay operator a number of ways are possible
to go beyond the current spectator model. For example, one could take into
account different quark-meson couplings and/or extended meson wave functions. 

\begin{acknowledgments}
This work was supported by the Austrian Science Fund (FWF-Projects
P16945 and P19035). B.~S. acknowledges support by the Doctoral
Program 'Hadrons in Vacuum, Nuclei and Stars' (FWF-Project W1203).
The authors have profited from valuable discussions with L.~Canton
and A.~Krassnigg.
\end{acknowledgments}

\end{document}